# Effect of Interdigital Capacitor on CRLH Leaky Wave Antenna Based on J-Shaped Metamaterial

Saeid Mohammadpour Jaghargh
Department of Telecommunication, Faculty of Electrical Engineering, Semnan University, Semnan, Iran
saeid.mohammadpour@semnan.ac.ir

ORCID 0000-0003-4734-7899

*Abstract—* **This paper aims to present a miniaturized novel reconfigurable composite right/left-handed leaky wave antenna (CRLH LWA) based on metamaterial as well as slow wave structure. In other words, the effect of interdigital capacitor (IDC) on LWA with periodic J-shaped metamaterial is investigated. This microstrip antenna is designed by cascading J-shaped metallic and IDC unit cells. Then it is simulated by full-wave ADS Momentum software. Advantageously, the proposed structure has tunable dispersion diagram that can be tuned by the number of IDC fingers. As will be clarified, altering the number of IDC fingers is able to adapt the equivalent capacitor of the antenna equivalent circuit. Thanks to the availability of RF switches technologies, changing the configuration of the IDCs can be practical. Furthermore, scanning the space from backward to forward region through broadside is one of the distinguishing features of the antenna. Indeed, the CRLH LWA continuously scans the space through broadside from -55° to +65° in $\varphi = 0°$ plane. The results of the investigation reveal that the LWA with two fingers, as more effective configuration, is a circularly polarized CRLH LWA and it is one of the highly desirable advantages of this structure. The axial ratio in the direction of the main beam for the backward region is lower than 3dB and it is improved in comparison to other regions. The deployment of IDC with variable figures in LWA with J-shaped metamaterial not only brings about the aforementioned significant features but also plays a pioneering role in this paper.**

## I. Introduction

Leaky wave antenna (LWA) is a guiding structure, supporting wave propagation along the length of the structure, with the wave radiating or leaking continuously along it. This microwave structure supports traveling waves as an antenna. The physics and operating rules of LWA as well as its instruction are summarized in [1]–[5]. The composite right/left-handed (CRLH) LWA, metamaterial inspiration, is one of the classifications of LWAs that is part of the periodic LWAs. Periodic LWA is periodically modulated in order to turn the non-radiating mode, surface wave, into a radiating leaky mode. CRLH LWAs beneficially have the ability to scan the radiation beam from backfire to endfire through broadside radiation whereas the conventional LWAs encounter a difficulty in achieving broadside radiation. This issue is known as open stopband problem. This poses travelling waves turn into standing waves while attenuation constant drops to zero and it is a deviation from the normal procedure. In addition, at this frequency interval the antenna suffers from mismatch at input and beam width of the antenna significantly falls, depending on length of the antenna. Dealing with this issue involves using metamaterial as left-handed (LH) in conventional LWA. Owing to using metamaterial, both LH and right-handed (RH) frequency regions are observed [1]-[2]. In the LH region, phase and group velocity are anti-parallel since effective refraction index (*n*), permittivity (*ε*), permeability (*μ*), and radiating wavenumber are negative although in the RH region, they are naturally parallel [6]. Consequently, at transition point, where radiating wavenumber is zero, the broadside radiation could be achieved. Wholly CRLH LWA has been reviewed in [5]. Beam scanning capability, high directivity, supporting both guided wave (slow wave) and radiating wave (fast wave) simultaneously, simple structure, low price, easy fabrication, low profile, millimeter wave (including 5G) and terahertz wave applications, beam forming applications ,and not complicated feed network (opposite of phased array antennas) are the desirable and attractive properties of CRLH LWA. Recent attention has been devoted to CRLH LWA due to accelerating level of interest and pace of development in metamaterial. Scientists and RF engineers latterly have concentrated on structures that are able to scan to endfire, structures that can scan through broadside (overcoming open stopband issue), conformal LWAs, directive optical LWAs, millimeter waves LWAs, terahertz LWAs ,and active beam shaping LWAs for beam forming [1]-[10].

For instance; a novel supercell based dielectric grating dual-beam leaky-wave antenna for WiGig (millimeter wave) applications is reported in [11]. This antenna suffers from open stopband issue; therefore, this LWA is not able to scan the space through broadside continuously. This structure and [12] also possess linear polarization. Unlike linear polarized (LP) antennas, circular polarized (CP) antennas are more popular in wireless systems, especially in satellite and mobile communication systems as well as global navigation satellite systems (GNSS). CP antennas in comparison to LP antennas possess a large number of great merits, including reducing the Faraday rotation effect, no strict orientation between transmitting and receiving antennas, and combating multi-path interferences [13]. The next antenna is a full-space scanning half-mode substrate integrated waveguide (HMSIW) LWA with ramp-shaped interdigital capacitor (IDC) that is studied in [14]. This structure is CP however it is neither tunable nor physically efficient in volume size in comparison to this work. Probably SIWs are not only as easy fabrication as planar microstrip technologies, due to presence of vias, but also have losses in dielectric substrate; as result, they could not be ideal for higher frequencies particularly in millimeter wave applications. Generally, tunable dispersion diagram causes that the antenna designer is able to select the accurate frequency scanning band in LWA applications without encountering complicated mathematical approaches. Another example is a programmable LWA with periodic J-shaped metamaterial patches [15]. This programmable LWA is able to tune the dispersion diagram in the direction of main beam. Full space scanning is one of its key features nevertheless its polarization was not noted. The last one is a micro-LWA that is proposed in [10] for terahertz radar application. Considering attractive properties of terahertz waves, including high spatial resolution and penetrability into nonmetallic materials, the combination of terahertz waves and LWA is responsible for a low-loss, low-profile, and wide-aperture terahertz radar. Nonetheless, [10] is unable to tune the dispersion curve.

In this paper, a microstrip CRLH semi-CP LWA based on J-shaped metamaterial and IDC unit cell (as slow wave structure) is proposed. The proposed antenna is remarkably accompanied by tunable dispersion diagram as well as full-space dynamic scanning capability. This microstrip LWA (MLWA) operates in X-band. As you are aware, the ability to operate in X-band frequency is highly desirable for radar applications. The proposed miniaturized CRLH LWA is designed and simulated by cascading 8-cells as that above-mentioned. Each cell is formed by a pair of metallic J-shaped metamaterials and two conventional IDCs with variable fingers. According to our previous researches [16]-[18], various and sundry of slow wave structures, and unlike [15], IDCs have been efficiently utilized in this work as series gap capacitors. Applying IDCs can lead to a more compact LWA effectively. Changing the number of IDC fingers is able to tune the LWA dispersion diagram (or broadside frequency) from ~9.3GHz to ~10.7GHz; however it has the adverse effects on polarization. The proposed antenna entirely and dynamically scans the space through transition frequency from -55˚ to +65˚ in $\varphi = 0˚$ plane without complicated feed network as consistently used in a phased array antennas.

## II. ANTENNA STRUCTURE, DESIGN, AND SIMULATION

In this section, the structure of proposed CRLH LWA based on a J-shaped metamaterial and IDC unit cell as slow wave structure will be presented. Then it will be simulated by the commercial Agilent-ADS Momentum software. The LWA with three different configurations will be simulated and tuning the dispersion diagram will be demonstrated. In fact, thanks to the availability of RF switches technologies, changing the configuration of the structure, i.e. altering the number of fingers, can be practical. As the first configuration, when the all RF switches are in an off state the IDC only has 2 active fingers, but, by contrast, the IDC has 4 active fingers when all switches are on. In each unit cell, turning off the upper switches can cause the IDC to play an active role with only 3 fingers. It is assumed that RF switches are PIN diodes. One of the most compelling reasons for using PIN diode switches are that they have shown to approximate, to a very good degree, ideal switches. Fig. 1 depicts layout and 3D view of CRLH LWA based on J-shaped metamaterial and IDC unit cell. The value of the fused quartz substrate thickness is selected to be $h = 1.5748$ mm in order to produce the desired leakage radiation. The relative permittivity of the substrate is $\varepsilon r = 3.8$. The mutual gap between the cells is 0.1mm as illustrated in Fig. 1.

An IDC is not only used as a simple, affordable, and high $Q$ slow wave structure but also is widely used as a quasi-lumped element in microwave integrated circuits (MICs) as well as high-speed integrated circuits. Hence, the presented structure is miniaturized and it is low profile.

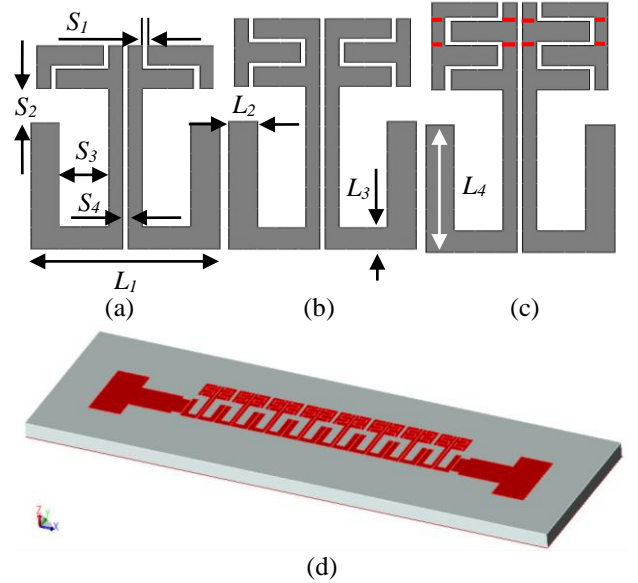

Fig. 1. Proposed CRLH circularly polarized LWA based on a J-shaped metamaterial and IDC unit cell. Dimensions of LWA cells with 2, 3 and 4 fingers are equal (except the number of IDC fingers). (a) The unit cell of proposed LWA. IDC only has 2 active fingers. Gray parts are metals. Each unit cell has a pair of J-shaped patches and it is symmetric. Parameter values are $L_1 = 3.1$ mm, $S_1 = S_4 = 0.1$ mm, $S_2 = 0.74$ mm and $S_3 = 0.81$ mm. Finger length and finger width of IDC are 0.86 mm and 0.4 mm respectively. (b) IDC has 3 active fingers. Parameter values are $L_2 = 0.47$ mm and $L_3 = 0.45$ mm. (c) IDC has 4 fingers. $L_4 = 2.66$ mm. The red parts are ideal RF switches. (d) 3D view of proposed LWA with eight periodic unit cells. Here, the red parts are metals that are located on the gray substrate.

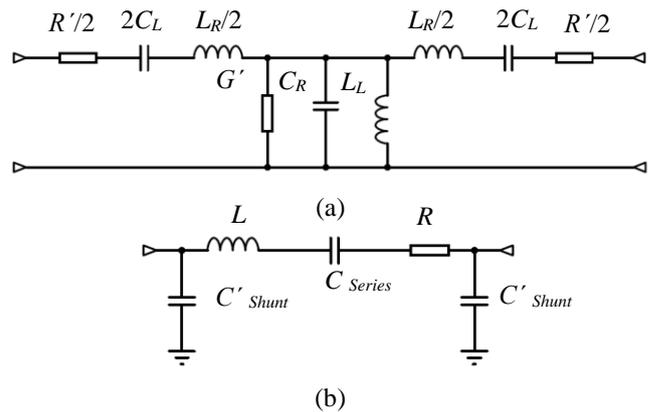

Fig. 2. (a) Lumped-element equivalent circuit model of proposed CRLH LWA unit cell. (b) Lumped-element equivalent circuit model of IDC.

Owing to utilizing slow wave structure (i.e., IDC) the unit cell dimensions of the presented antenna developmentally become more optimized in comparison to [15]. Slow wave structures are popular in circuit miniaturization, frequency-selective devices and filters, travelling wave devices, time-delay lines, phase shift and equalization, impedance match, and pulse shaping applications. Across the narrow, long, and folded gap between the two thin-film conductors that is used in IDC, series capacitance intrinsically occurs. In other words, most of capacitance can be accessed at the edge of fingers. Hence, this long length of the gap makes the unit cell smaller [19]-[20]. The equivalent circuit of proposed CRLH LWA is reported in [15] but the lumped circuit does not include the slow wave structure that is used in the unit cell. Fig. 2 illustrates both equivalent circuits. Equivalent circuit model of an IDC is effortlessly capable of revealing slow wave propagation of a mode with the characteristic lumped elements per periodic cell, denoted by $RCLC'$. Longitudinal current flow on the lossy conductor, the longitudinal current flow, and impact of transverse electric field produce $R$, $L$, and $C'$ respectively. According to one of the essential design considerations the dimensions of the capacitor are required to be smaller than wavelength in order to play an active role as a lumped element. Expressions for these elements are subsequently noted. The period of the structure is $p = 3.2$ mm and the series capacitance of a single-layer IDC is able to be approximated as [20]-[22]:

$$C_{Series} = (l/w) \times (\varepsilon_r + 1) \times [(N-3)A_1 + A_2] \quad (1)$$

Where $N$, $l$, $w$, $A_1 = 0.089$ pF/cm and $A_2 = 0.1$ pF/cm are the number of fingers, finger length, finger-base width in cm, contribution of the interior finger for $h > w$, and the contribution of the two external fingers for $h > w$ respectively. It is fairly obvious that the amount of series capacitance of IDCs cannot be only risen by increasing the number of fingers but also can be increased by using a thin layer of high dielectric constant material, for example ferroelectric, between the conductors and the substrate alternatively. Furthermore, the length of fingers is directly proportional to the $C_{Series}$. Series gap capacitance can also be calculated as [20]:

$$C_{Series} = \frac{10^{-3}\varepsilon_r}{18\pi} \times \frac{K(k)}{K'(k)} \times (N-1) \times l \quad (pF) \quad (2)$$

Where $l$ is finger length in micrometer, and the ratio of complete elliptic integral of first kind $K(k)$ and its complement $K'(k)$ are given by (3):

$$\frac{K(k)}{K'(k)} = \begin{cases} \frac{1}{\pi}\left[\ln\left(2\frac{1+\sqrt{k}}{1-\sqrt{k}}\right)\right] & ; 0.707 \leq k \leq 1 \\ \frac{\pi}{\ln\left(2\frac{1+\sqrt{k'}}{1-\sqrt{k'}}\right)} & ; 0 \leq k \leq 0.707 \end{cases} \quad (3)$$

Also $k$, $k'$, $a$, and $b$ can be expressed as:

$$k = \tan^2\left(\frac{a\pi}{4b}\right) \Rightarrow k' = \sqrt{1-k^2} \quad (4)$$

$$a = \frac{W'}{2} \Rightarrow b = \frac{(W'+S)}{2} \quad (5)$$

where in this design $W' = S = 0.4$ mm is finger width. The series resistance of IDC is given by:

$$R = \frac{4}{3}\left(\frac{l}{W'N}\right) \times R_s \quad (6)$$

where $R_s$ is the sheet resistivity. Consequently, $C_{series} = 0.0867$ pF is calculated for an IDC with 4 fingers. Shunt capacitance $C'_{Shunt}$ and the series inductance $L$ of the IDC could be calculated from microstrip transmission-line theory as:

$$L = \left(\frac{Z_0\sqrt{\varepsilon_r}}{c}\right) \times l \quad (7)$$

$$C'_{Shunt} = \frac{1}{2}\left(\frac{\sqrt{\varepsilon_r}}{Z_0 c}\right) \times l \quad (8)$$

where $Z_0$, characteristic impedance, is calculated using $w$ and $h$ IDC parameter. Here, $c \approx 3.00 \times 10^{10}$ cm/s is the velocity of light in free space. As mentioned above, increasing the number of fingers, by turning the switches on, will rise the inductance $L$, capacitance $C'_{Shunt}$ as well as capacitance $C_{Series}$. By contrast, it will decline the resistance $R$ [20]. Also it has a beneficial effect on perturbing and redistributing current so that the capacitances, inductances in the equivalent circuit of proposed CRLH LWA, and phase constant $\beta$ are tuned [15]. The tuned phase constant $\beta$ produces the desired result, is named tunable dispersion diagram. Another effect is that the magnitudes of S-parameters witness noticeable resonances in high frequency bands due to this increasing, as indicated in Fig. 3. Thanks to availability of shunt inductors and series capacitors in the equivalent circuit of proposed LWA unit cell, LH elements, the antenna is claimed to be CRLH. This antenna is analyzed from 7.5 GHz to 12 GHz. Besides all aforementioned justifications for using this geometry in presented antenna, periodic LWAs can be classified as transversally or longitudinally symmetric or asymmetric. The presented geometry of the LWA is longitudinally asymmetric. Breaking of the symmetry of periodic LWA with respect to its longitudinal axis can necessarily bring about elliptical or circular polarization as well as overcoming broadside gain degradation in any periodic leaky-wave antenna [23]. Hence, this one may be a fundamental justification for presenting this structure geometry that will be accurately investigated in section III.

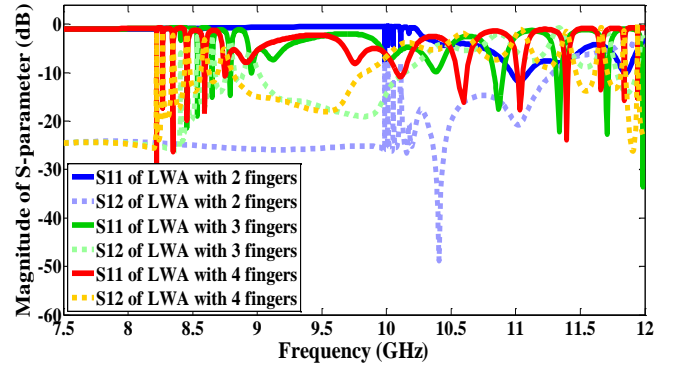

Fig. 3. Magnitudes of S-parameters for proposed CRLH LWA.

III. THEORY, COMPARISON, AND RESULTS

Probably conventional wireless telecommunications systems with a fixed frequency band and fixed radiation pattern as well as linear polarization could practically become outdated due to expanding the network in revolutionary applications, e.g. internet of things (IoT), 5G technology, smart cities, and healthcare [24]. In direct contrast to traditional antenna, CRLH LWA is intrinsically capable of altering the radiation pattern without utilizing phase shifters.

The presented antenna is a periodic structure. Periodicity is responsible for spatial separation of electric and magnetic energy or increasing the reactance (inductance, capacitance, or both) in antenna equivalent circuit. The separation of energies is a fundamental condition for propagating the slow wave structures. This physical property can be formulated by a comprehensive theory. Periodicity or Floquet's theorem not only plays roles of paramount importance in formulating a linear, periodic and, slow wave system but also plays a significant role in determining CRLH LWA properties, including slow wave, fast wave, leaky wave (LW) mode, LH, and RH region. The accurate determination of periodic CRLH LWA phase constant $\beta$ at a specific frequency $f$ is the principal issue in Floquet's theorem. It is maintained that the dispersion diagram is the way forward [19]. There are three approaches that could be adopted for extracting dispersion curve in CRLH LWA unit cell. The first one is Eigen-mode simulation using periodic boundary conditions with Ansys products [25]. Another approach is to extract dispersion diagram from S-parameters [14], [26]. Furthermore, calculating dispersion diagram could be addressed by utilizing the following formula. According to formula 9, tunable dispersion curve of the proposed CRLH LWA (with 2, 3, or 4 fingers) is extracted from scan angles in which the transition frequency from the LH to the RH shifts from 9.3 GHz to ~10.7 GHz, as the number of IDC fingers increased or declined. It is illustrated in Fig. 4. Radiating mode (or fast wave), guiding mode (or slow wave), LW mode, RH radiation, and LH radiation of each antenna too is depicted in Fig. 4.

$$\theta_s = Sin^{-1}\left(\beta/k_0\right) \quad (9)$$

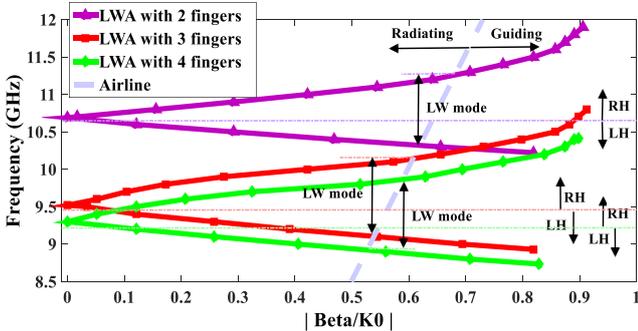
Fig. 4. Tunable dispersion curve of the proposed CRLH LWA unit cell due to changing the number of IDC fingers, extracted from scan angles.

Moreover, scan angles or radiation direction $\theta_s$ of LWA can be obtained from formula 9 where $\beta$ and $k_0$ are the phase constant and the wave number in the air respectively. The dispersion diagrams have confirmed that the proposed CRLH LWA never encounters stopband problem owing to absence of bandgap region between the LH and RH regions. Thus the presented CRLH LWA continuously is able to scan the space from backward to forward region through broadside. In other words, the proposed antenna is a balanced antenna with the balancing points (or broadside radiation) located at about 10.7, 9.5, and 9.3 GHz for 2, 3, and 4 fingers respectively. Balanced case absolutely happens when the series resonance frequency, shunt resonance frequency, and broadside frequency are equal. Hence $\omega_{se}$ and $\omega_{sh}$ are given by:

$$\omega_{se} = \frac{1}{\sqrt{L_R C_L}} \quad and \quad \omega_{sh} = \frac{1}{\sqrt{L_L C_R}} \quad (10)$$

where $L_R$, $C_L$, $L_L$, and $C_R$ are the equivalent circuit elements [1]-[5]. It is also validated that altering the number of IDC fingers or changing the gap state where the length of unit cell is smaller than one-fifth of the wave length could lead to a tunable dispersion diagram. The presented CRLH LWA with tunable dispersion diagram is straightforwardly able to be designed in a specific frequency band without encountering complicated mathematical approaches.

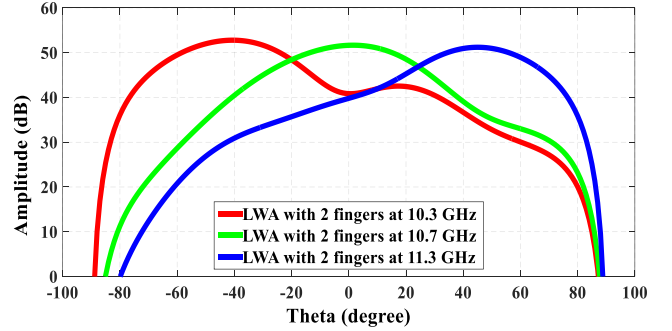

(a)

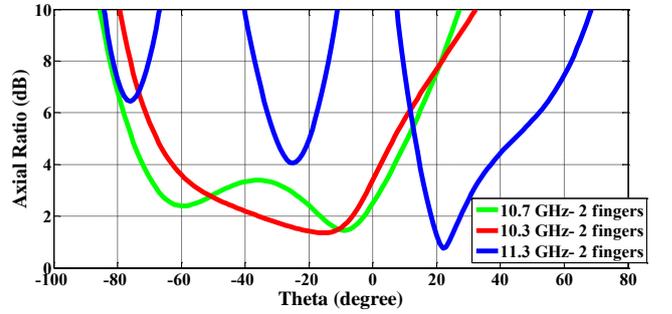

(b)

Fig. 5. (a) Radiation pattern of proposed CRLH LWA with two IDC fingers and scan angles. (b) The axial ratio in the direction of the main beam at 10.3, 10.7, and 11.3 GHz.

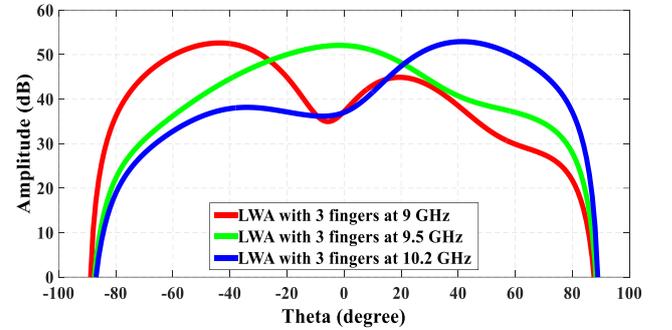

(a)

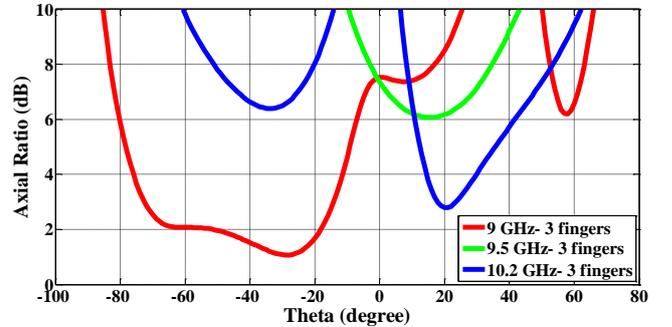

(b)

Fig. 6. (a) Radiation pattern of proposed CRLH LWA with three IDC fingers and scan angles. (b) The axial ratio in the direction of the main beam at 9, 9.5 and 10.2 GHz.

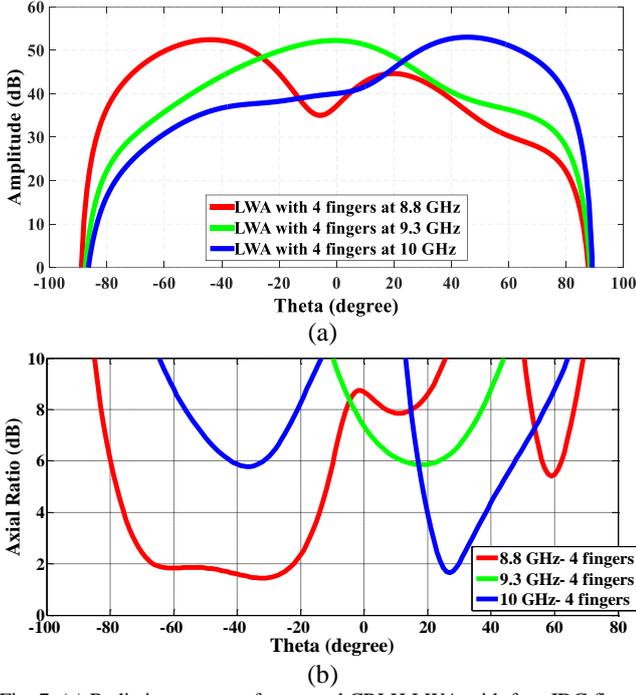

Fig. 7. (a) Radiation pattern of proposed CRLH LWA with four IDC fingers and scan angles. (b) The axial ratio in the direction of the main beam at 8.8, 9.3 and 10 GHz.

As mentioned above, the main lobe of the radiation pattern is continuously capable of scanning the space through broadside when frequency changes. Main beam of proposed LWA continuously scans from $\theta_1 = -55°$ to $\theta_2 = +65°$. The scanning operation has been investigated in $\varphi = 0°$ plane. Fig. 5(a), 6(a), and 7(a) demonstrate the radiation patterns and scan angles of proposed CRLH LWA. Some frequencies have been selected in order to exhibit the continuous scanning.

Details of the proposed LWA scan angles are demonstrated in Fig. 8. Furthermore, the axial ratio, a dispensable parameter for determining the circular polarization, in the direction of the main beam is illustrated in Fig. 5(b), 6(b), and 7(b). Axial ratio is usually required to be below 3 dB for a CP antenna. CP antennas are more popular owing to reducing the Faraday rotation effect, no strict orientation between transmitting and receiving antennas, and combating multi-path interferences as that above-mentioned [13]. Due to this popularity, recent attention has been even devoted to polarization converters. These structures convert linear polarized signals to circular ones [27]-[28]. In this work, the observations reveal, as illustrated in Fig. 9, that the axial ratio in the direction of main beam for backward region is superior to other regions especially for configuration with 2 IDC fingers. Surprisingly, polarization of it is lower than ~3 dB in backward region.

Conventional LWAs suffer from gain degradation within the pattern bandwidth. Metamaterial, specifically J-shaped metamaterial in this work, is responsible for enhancing the gain and reducing the gain variation [9]. Hence metamaterial, classification as a periodic LWA as well as longitudinally asymmetry can lead to avoid gain degradation in the broadside direction due in part to equality of series and shunt efficiencies [23]. While tuning or scanning, variations of gain for presented antenna with 2 IDC fingers are fairly low. As you are aware now, variation of the gain in LWA is a significant parameter that has been examined below. Notice Fig. 10.

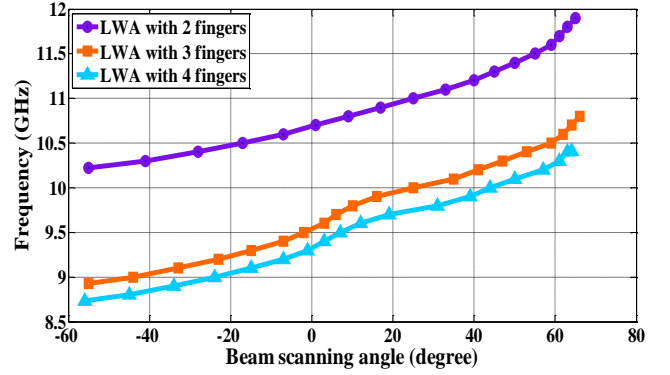

Fig. 8. Details of beam scanning angles in proposed CRLH LWA when the frequency changes.

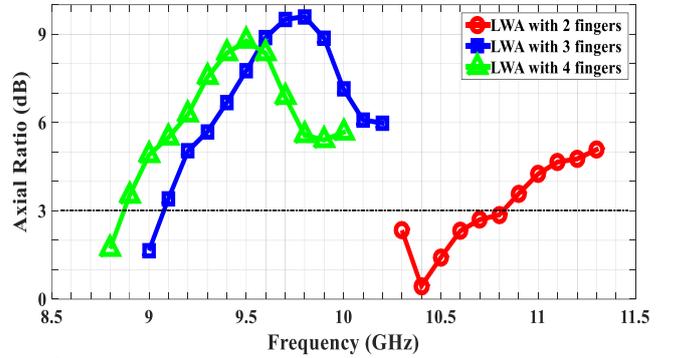

Fig. 9. Details of the axial ratio in the direction of the main beam in proposed CRLH LWA with 8-cells when the frequency changes.

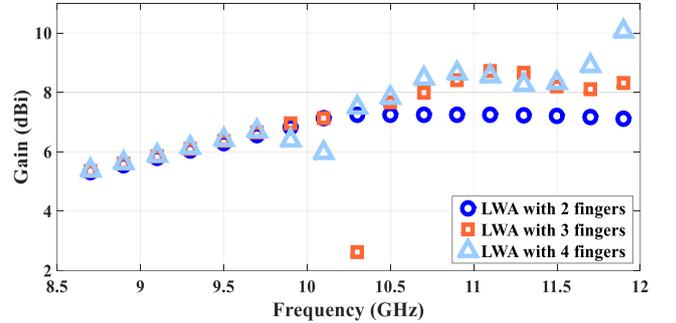

Fig. 10. Details of magnitude of the gain in proposed CRLH LWA with 8-cells when the frequency changes.

The proposed CRLH MLWA is compared with some of the prior works in the literature in Table I. Unlike [11], due to using the metamaterial and longitudinally asymmetric unit cell, the proposed CRLH LWA is capable of scanning the space without encountering the stopband issue. Furthermore, owing to utilizing the IDCs, the presented CRLH LWA possesses tunable dispersion curve in comparison with [10] to [14]. Considering frequency band, miniaturized dimensions as well as circular polarization, notably for structure with 2 IDC fingers, are another significant attainment of this work.

TABLE I. COMPARISON OF THE MINIATURIZED CRLH MLWA WITH OTHER RECENT WORKS

| Structure | Antenna Type | Backward & Forward Scanning / Broadside Radiation | Open Stopband Problem | Tunable Dispersion Curve | Freq. Band | Pol. | Unit Cell Dimensions (mm) | Max Gain (dBi) |
|---|---|---|---|---|---|---|---|---|
| [10] | MLWA for THz applications | Yes / Yes | No | No | THz | N/A | 0.73 * 0.8 | ~19 dB |
| [11] | Dielectric Grating Dual-beam LWA | Yes / No | Yes | No | V | LP | 0.97 * 8 | ~25 |
| [12] | CRLH MLWA | Yes / Yes | No | No | C | LP | 6 * ~19 | ~15 |
| [14] | HMSIW with ramp-shaped IDCs | Yes / Yes | No | No | X | CP | 12 * 10 | 12.05 |
| [15] | MLWA with J-shaped MTM | Yes / Yes | No | Yes | X | N/A | 3.1 * 5.19 | N/A |
| This work | MLWA with J-shaped MTM and IDCs | Yes / Yes | No | Yes | X | ~ CP | 3.1 * 4.3 | ~10 |

## IV. CONCLUSIONS

The author has exhibited a novel miniaturized microstrip CRLH LWA based on metamaterial and slow wave structure with tunable dispersion. A pair of metallic J-shaped metamaterials and a pair of IDCs with variable fingers constitute the unit cell. Broadside of the LWA can be tuned from ~9.3GHz to ~10.7GHz by the number of IDC fingers. Proposed LWA, the examinations showed, continuously scans the space through broadside from -55° to +65° in φ=0° plane. Additionally, effects of slow wave structure, i.e. IDC, have been investigated. Investigations demonstrated that the axial ratio of the structure with 2 IDC fingers, as the best configuration, in the direction of the main beam for the backward region is lower than ~3 dB.

## ACKNOWLEDGEMENTS

The author would like to acknowledge Semnan University for its assistance. Moreover, the author profoundly expresses his gratitude for editing services that Faezeh Mohammadpour Jaghargh has provided.